\documentclass[article]{revtex4}
\usepackage{epsfig}
\newcommand{\beq}{\begin{eqnarray}}
\newcommand{\eeq}{\end{eqnarray}}
\begin{document}
\title{ Renormalization group and bound states 
\footnote[1]{Presented at LIGHT CONE 2008 Relativistic Nuclear and Particle Physics, July 7-11, 2008, Mulhouse, France} }
\author{ Stanis{\l}aw D. G{\l}azek } 
\affiliation{Institute of Theoretical Physics, University of Warsaw, Poland}
\begin{abstract} 
Similarity renormalization group procedure
identifies the role of bound states in the low-energy
rate of change of effective coupling constant in a
model Hamiltonian with asymptotic freedom.
\end{abstract}
\pacs{12.38.-t,12.39.-x,12.90.+b,11.15.-q}
\maketitle

\section{Introduction}
\label{sec:I}
The fact that an asymptotically free interaction
blows up at low energies is often associated
with the concept of confinement. It is therefore
of interest to construct a soluble model that
exhibits a similar variation of a coupling
constant and see what mechanism causes that the
coupling constant rises at low energies.

In a very simple model with asymptotic freedom and
a bound state, similarity renormalization group
(SRG) procedure for Hamiltonians produces a strong
effective interaction at low energies. The
interaction strength is related to the bound-state
formation and a degree to which the interactions
that are responsible for the binding are included
in the generator of SRG transformations: the more
interaction in the generator the less increase in
the coupling constant~\cite{SRGf}. Similar results 
are found in the case with limit cycle, of which 
the apparent asymptotically free SRG behavior of 
effective interactions may be a part. This lecture 
concerns only some aspects of these results. Readers 
interested in a bit broader picture may consult 
Ref.~\cite{RGandBSAPP}. A connection with AdS/CFT
correspondence~\cite{Maldacena,MaldacenaReview}
that inspires a new generation of models for
calculating masses of bound states of asymptotically 
free quarks and gluons ~\cite{BrodskyTeramond1,BrodskyTeramond2},
is not yet established.
\section{Model}
\label{sec:Model}
Consider a model Hamiltonian in the generic form of
\beq
H & = & H_0 + H_I \, .
\eeq
$H_0$ can be just a free (kinetic) energy or include
also dynamical effects that are well understood. The 
origin of division of $H$ into $H_0$ and $H_I$ is that 
all eigenvalues and eigenstates of $H_0$ are known. 
Thus, $H_0$ provides a basis for studying new effects 
due to the postulated $H_I$. For simplicity of the 
model, suppose the eigenvalue problem for $H_0$,
\beq
H_0 |n \rangle & = & E_n |n \rangle \, ,
\eeq
is solved with a set of discrete eigenvalues $E_n > 0$ 
that form a monotonic sequence, $E_m > E_n$ when $m > 
n$, there is no degeneracy, and the corresponding 
eigenvectors are normalized, $\langle m | n \rangle = 
\delta_{mn}$. The interaction Hamiltonian $H_I$ is 
then defined by its matrix elements $H_{Imn} = \langle 
m | H_I | n \rangle$. An analytically soluble model to
be discussed here, is obtained by assuming that these 
matrix elements have a factorized form, i.e., $H_{I mn} 
= H_m H_n$. $H_n$ should have dimension of square root 
of energy and the simplest choice that does not introduce 
any dimensionful scale is $H_n \sim \sqrt{E_n}$. The 
proportionality is reduced to a dimensionless number, 
and one can write 
\beq
\label{HI}
\langle m | H_I | n \rangle   = 
- g \sqrt{E_m} \sqrt{E_n} \, ,
\eeq
where $g$ determines the strength of the interaction,
called coupling constant in analogy with QFT. The
negative sign results in existence of a bound state 
for sufficiently large positive $g$. By definition, 
the bound state corresponds to a negative eigenvalue 
of $H$. It will have to be clarified what the words 
``sufficiently large $g$'' are supposed to mean, 
because the model Hamiltonian produces divergences 
(infinities) no matter how small the number $g$ is.

In summary, the model Hamiltonian is defined by its 
matrix elements as follows:
\beq
H_{mn} 
& = & 
\langle m | H | n \rangle  
  = 
E_m \delta_{mn} - g \sqrt{E_m} \sqrt{E_n}
\, .
\eeq
\section{ Ultraviolet divergences }
\label{sec:UV}
Suppose $g$ is a very small number and one expects 
that eigenvalues of $H$ should be nearly equal to
the eigenvalues of $H_0$. First-order correction 
to $E_m$ is a fraction $g$ of the energy being
corrected,
\beq
\Delta E_m^{(1)} 
& = & 
\langle m |H_I | m \rangle
  =
-g \, E_m \, .
\eeq
The second-order correction is 
\beq
\Delta E_m^{(2)} 
& = & 
\sum_{k \neq m} {  | \langle m |H_I|k \rangle |^2  \over E_m - E_k  } 
 =  
g^2 E_m \, 
\sum_{k \neq m} {E_k \over E_m - E_k} \, .
\label{2nd}
\eeq
Terms with $k < m$ involve $E_k < E_m$ and if $E_k
\leq E_m/b^{m-k}$ with a number $b > 1$, they
contribute an amount that is not sensitive to the
lower bound on $E_k$, say $b^M$ with a large
negative integer $M$, when $M \rightarrow
-\infty$. Terms with $k \gg m$ involve $E_k \gg
E_m$ and thus each of them contributes $-1$. If
$E_k = b^{k-m} E_m$, and $b \gg 1$, the
second-order correction to $E_m$ is proportional to 
the number of basis states with energies (eigenvalues 
of $H_0$) greater than $E_m$. In order to obtain 
a finite correction, one has to limit the number of 
the basis states and try to understand what happens 
when the limit is relaxed. In analogy with QFT, 
imposing such a limit is called regularization. If 
one assumes a cutoff of the form $k \le N$, where 
$N$ is a large positive integer, the result of 
Eq.~(\ref{2nd}) for $b \gg 1$ is
\beq
\label{N}
\Delta E_m^{(2)} 
& = & 
- g^2 \, (N - m) \, E_m \, .
\eeq
The ultraviolet cutoff on energies, $\Lambda = 
b^{N-m_0} E_{m_0}$ means $k \leq N = m_0 + 
{ 1 \over \ln b } \, \ln{\Lambda /E_{m_0}}$, and one says that
the second-order correction is ultraviolet divergent 
logarithmically,
\beq
\label{Lambda}
\Delta E_m^{(2)} 
& = & 
- g^2 \, {1 \over \ln b } \, \ln {\Lambda \over
E_m} \, \, E_m \, .
\eeq
The divergence results from all different energy 
scales contributing equally to the correction.
The occurrence of divergences is not limited to 
perturbation theory. 

With the factorized interaction, solutions to the 
eigenvalue problem 
\beq
\label{E}
\sum_{n = M}^N H_{mn} \psi_n & = & E \psi_m \, ,
\eeq
have wave functions
\beq
\label{psi}
\psi_m & = & { \sqrt{E_m} \, 
\over  E_m - E}\,\, g
\sum_{n = M}^N \sqrt{E_n} \psi_n 
\, ,
\eeq
in which the eigenvalue satisfies the condition
\beq
\label{binding}
1 + g \sum_{n = M}^N { E_n \over  E - E_n} & = & 0 \, .
\eeq
The sum here resembles closely the one in Eq. (\ref{2nd}) 
and develops the same type of divergence. If a
bound state exists, with a negative eigenvalue $E
= - E_B$, one can replace the sum over $n$ by an integral, 
$dn = dE_n/(E_n \ln b )$, to estimate what
happens, and obtains
\beq
\label{E_B}
E_B & = & { \Lambda - b^M e^{\ln b \over g }
\over  e^{\ln b \over g } - 1 } \, .
\eeq
In the limit of $\Lambda \rightarrow
\infty$ for fixed $g$, the binding energy diverges.
One can also observe that the square of the matrix
$H_I$ with large $N$ and $M$ is equal to ${g b
\Lambda \over 1 - b} \, H_I $, which means that it
diverges in the ultraviolet limit of $\Lambda
\rightarrow \infty$ for fixed $g$. This means that
all powers of the entire $H$ are ultraviolet
divergent. In particular, the evolution operator
$U(t,0) = e^{-iHt}$ does not exist in this limit. 

\section{ Asymptotic freedom }
\label{sec:AF}

The trouble with divergences can be summarized using 
the model as follows. One measures transition rates 
between, say, two states, say $|m_1 \rangle$ and $|m_2\rangle$, 
and discovers that these rates can be described in first-order 
perturbation theory by an interaction Hamiltonian with 
matrix elements
\beq
\label{2x2}
\left[
\begin{array}{ll}
\langle m_2 | H_I | m_2 \rangle , & \langle m_2 | H_I | m_1 \rangle \\
\langle m_1 | H_I | m_2 \rangle , & \langle m_1 | H_I | m_1 \rangle \\ 
\end{array}
\right]
& = & 
-g^{(2)}
\left[
\begin{array}{ll}
E_{m_2}                  , & \sqrt{ E_{m_2} E_{m_1} }  \\
\sqrt{ E_{m_1} E_{m_2} } , & E_{m_1}                   \\ 
\end{array}
\right] \, .
\eeq
The coupling constant $g^{(2)}$ corresponds to physics of the 
2 states. One is then compelled to postulate that the whole 
matrix of $H_I$ has the form given in Eq.~(\ref{HI}). This 
step is analogous to the proposal of non-Abelian gauge 
theory~\cite{YangMills}. Even if this leads to divergences, 
one does not want to abandon the proposed interaction since
it does produce a structure in perturbation theory that 
fits the case of states $|m_1 \rangle$ and $|m_2\rangle$ 
and it has an appealing symmetry. Therefore, one looks for a general 
way out of the problem with divergences that are produced by 
naive extrapolation of knowledge from a small set of matrix 
elements to a large set. The large set is desired when one 
seeks a theory of presumably large range of applicability 
and a lot of predictive power.

The way to proceed is to learn what happens when one begins 
with some large $N$ and tries to mathematically reduce $N$ 
to a small value near $m_2$. As a principle, such procedure 
was proposed and developed in seminal Refs.~\cite{Wilson:1965,
Wilson:1970}. It is sometimes called ``integrating out 
high-energy degrees of freedom,'' which applies also in 
statistical mechanics~\cite{WilsonRG}. In the model, one 
starts with the eigenvalue problem 
\beq
H |\psi \rangle & = & E |\psi \rangle \, , 
\quad \quad \quad 
|\psi \rangle     =  \sum_{k = M}^N \psi_k
|k\rangle \, ,
\eeq
and applies Gaussian elimination, beginning with $\psi_N$. 
The remaining set of equations for $\psi_k$ with $k \leq N-1$ 
corresponds to (the required algebra is merely solving
one linear equation)
\beq 
H^{(N-1)}_{I mn}
& = & 
- \left( g - { g^2 E_N \over E - E_N + g E_N} \right)
\,
\sqrt{ E_m E_n } \, .
\label{H1model}
\eeq
$H_I^{(N-1)}$ has the same structure of matrix
elements as $H_I$ but contains a new ``coupling
constant'' (the expression in the bracket). 
A simplification occurs for cutoffs much larger 
than $|E|$, for which $E/E_N$ can be neglected,
\beq 
g^{(N-1)}
& = & 
g \, { 1 - E/E_N \over 1 - g - E/E_N } 
\sim 
\, { g \over 1 - g }\, .
\label{g1model}
\eeq
Therefore, one obtains a recursion that does not 
depend on the eigenvalue $E$,
\beq 
g^{(K-1)}
& = & 
{ g^{(K)} \over 1 - g^{(K)} } \, ,
\label{recursion}
\eeq
for as long as $\Lambda_K = b^K \gg |E|$.
This is the RG recursion in the model. It is solved by
\beq 
g^{(K)}
& = & 
{ g \over 1 - g (N-K) } \, . 
\label{solution1}
\eeq
Suppose the eigenvalue problem for $H^{(K_0)}$ with 
$b^{K_0} = \lambda_0$ is small enough to solve 
for its spectrum precisely using computers and
establish that the coupling constant $g^{(K_0)}$ 
should have some value $g_0$ in order to
reproduce some measured eigenvalue $E_0$ with 
$|E_0| \ll \lambda_0$ (some transition amplitude 
could be used instead). For $E_n = b^n$, 
Eq.~(\ref{solution1}) says that 
\beq 
g_0
& = & 
{ g_\Lambda \over 1 - {g_\Lambda \over \ln b} \ln
\Lambda/\lambda_0 } \, ,
\quad \quad 
g_\Lambda
 = 
{ g_0 \over 1 + {g_0 \over \ln b} \ln
\Lambda/\lambda_0 } \, .
\eeq
This means that the model is asymptotically
free~\cite{af1,af2}: the larger the cutoff
$\Lambda$ in the initial $H$ the smaller the
coupling constant $g_\Lambda$ in it, and the
smaller value is required to describe high-energy
processes in perturbation theory. Calculation of
$g_\Lambda$ means also evaluation of the counterterm 
required in the initial Hamiltonian.

The increase of $g^{(K)}$ when $K$ decreases causes 
a major difficulty because the ratio $E/E_K$ is 
compared with $1 - g^{(K)}$ in the RG recursion
and the unknown eigenvalue $E$ cannot be ignored 
when $g^{(K)} \rightarrow 1$ no matter how much
smaller $|E|$ is than $E_K$. This is a basic obstacle
to description of bound states in theories with 
asymptotic freedom: the eigenvalues cannot be 
found in perturbation theory and one cannot easily 
reduce the cutoff to sufficiently low values for 
carrying out all kinds of interesting non-perturbative
calculations using available computers. In order to 
control what happens in the range where $g^{(K)}$ 
is order 1, a different RG procedure seems appropriate.
 
\section{ Similarity RG procedure and $g_\lambda$ at low energies }
\label{sec:SRG}

In the SRG procedure~\cite{SRG}, one proceeds according 
to similar principles as in the standard approach
described in the previous section. One also finds 
counterterms and evaluates effective Hamiltonians. 
The new idea is that one does not ``integrate out'' 
any degrees of freedom. Instead, one changes the 
basis states by rotating them in the Hilbert space. 
The rotation is designed in such a way that it guarantees 
the resulting Hamiltonian matrix, $H_\lambda$, to have 
vanishing matrix elements between basis states if they
differ in energy by more than $\lambda$. The design is 
such that one does not encounter small energy denominators 
even when the coupling constant increases to 1 or larger 
values. Of course, the SRG procedure produces the same 
$CT$ as the one identified in the previous section and 
the initial condition at $\lambda = \infty$ for the SRG 
evolution of $H_\lambda$ with $\lambda$ is the same
\beq
\label{IC} 
H_{\Lambda m n} & = & 
E_m \delta_{mn} - g_\Lambda \sqrt{E_m} \sqrt{E_n}
\, .
\end{eqnarray}

The SRG evolution can be obtained from (prime denotes 
differentiation with respect to $\lambda$)
\beq
\label{SRGH}
H'_\lambda 
& = & 
[T_\lambda, H_\lambda ] \, , 
\eeq
where $T_\lambda = [G_\lambda, H_\lambda]$, $ G_\lambda 
= fH_0 + (1-f)D_\lambda$, and $D_\lambda$ denotes the 
diagonal part of $H_\lambda$. For $f=0$, one has 
$G_\lambda=D_\lambda$, in which the diagonal part of 
interactions is fully included and Eq.~(\ref{SRGH}) 
is the one introduced by Wegner in condensed matter
physics~\cite{Wegner1,Wegner2}. For $f=1$, one has 
$G_\lambda=H_0$, in which no interaction effects are included, 
and Eq.~(\ref{SRGH}) is then the one used in nuclear 
physics~\cite{PerrySzpigel,Bogner:2006srg,Bogner:2007jb}.
For intermediate values of $f \in [0,1]$, $G_\lambda$ 
includes interactions to an intermediate degree, 
correspondingly, and one can inspect what happens 
in various cases. 

\begin{figure}[ht] 
\begin{center}
\includegraphics[scale=0.6]{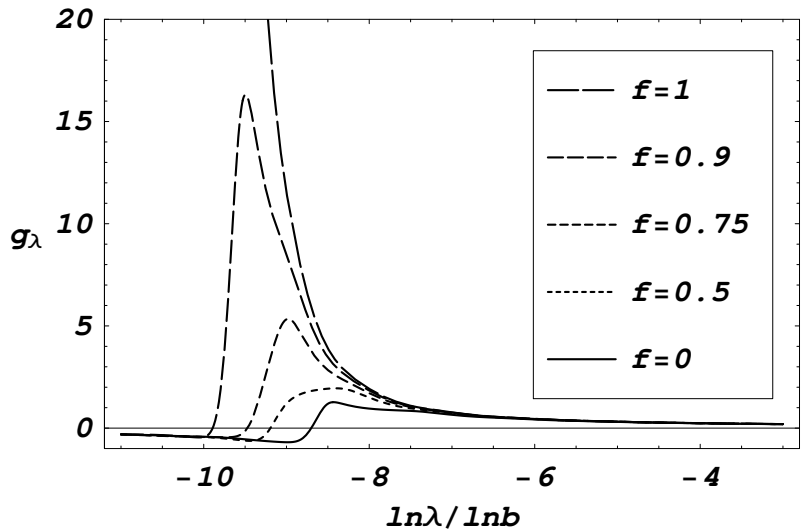} 
\end{center}
\caption{ \label{fig:gaf} {\small $g_\lambda$ 
grows toward small $\lambda$ differently for 
$f=0$, 0.5, 0.75, 0.9, and 1 (the larger $f$,
the higher curve). The huge increase of
$g_\lambda$ below $\ln{\lambda}/ \ln{b}
\sim -8$ for $f=1$, occurs because $\lambda$ 
decreases there below the scale of binding energy 
$E_B$. In fact, $g_\lambda \rightarrow |E_B|/b^M 
\sim 10^{10}$. For $f=0$, the huge increase of 
$g_\lambda$ is absent.}}
\end{figure}

Numerical calculations produce results that can be
summarized by writing 
\beq
\label{Hlambda}
H_{\lambda mn} 
 & \sim & 
\left[ E_m \delta_{mn} - g_\lambda 
\sqrt{ E_m } \sqrt{ E_n} \right] e^{-(E_m -
E_n)^2/\lambda^2} \, .
\eeq
The effective coupling constant $g_\lambda$ is then 
defined using the interaction matrix elements between 
the states of lowest energies, such as $|m_1\rangle$ 
and $|m_2\rangle $ in Eq.~(\ref{2x2}), in analogy with 
the Thomson limit in QED, say $m_1 = M$, $m_2 = M+1$,
and 
\begin{eqnarray} 
\label{gdefinition}
g_\lambda & = & 1 -
H_{\lambda M M+1}/\sqrt{E_M E_{M+1} } \, . 
\end{eqnarray}
A generic example is shown in Fig.~\ref{fig:gaf}, where
$b = 4$, $N=16$, $M=-25$,
$g_{\lambda =\infty} = g_\Lambda \sim 4/100$, and
$\Lambda = 4^{16} \sim 4 \cdot 10^9$. The
bound-state energy is $E_B \simeq -8 \cdot 10^{-6}$.
Fig.~\ref{fig:gaf} demonstrates that the increase 
of $g_\lambda$ at small energies is caused by
removing interactions responsible for existence of 
a bound state from $G_\lambda$ in the generator of 
the SRG transformations. When the generator fully 
accounts for the interactions, the magnitude of the 
coupling constant $g_\lambda$ never significantly 
exceeds 1. 

An apparently very small alteration of $H_{Imn}$
in the model, by a term $-i \,h \,{\rm sgn}(m-n) \sqrt{
E_m E_n}$ with a very small coupling constant $h$,
leads to a new way of thinking about asymptotic
freedom as a part of a limit cycle. RG limit
cycles were introduced in the context of strong
interactions in Ref.~\cite{WilsonGML} and recently
suggested relevant to the infrared behavior of
QCD~\cite{Braaten}. If $h$ is an irrational
number, the model typically exhibits a chaotic RG
behavior. When $h = \tan{\pi\over p}$ with $p$ a
large integer, a limit cycle occurs, with a period
$b^p \sim e^{\pi/h}$. This means that
$g_{\lambda_1}$ has the same value as
$g_{\lambda_2}$ if $\lambda_1 = (b^p)^k \lambda_2$
with integer $k$. The cycle is associated with
existence of bound states whose binding energies
form a geometric series with quotient $1/b^p$. If
a RG cycle were indeed present in a realistic
extension of the standard model with some tiny
coupling constant like $h$, say $\kappa$, a new
generation of particle substructure would be
predicted with binding energies order
$e^{\pi/\kappa} \Lambda_{QCD}$. 

\section{ Conclusion} 
\label{sec:C}
The simple model shows that the SRG procedure may be 
a suitable tool to handle the increase of the coupling 
constant $g_\lambda$ in QCD when $\lambda \rightarrow 
\Lambda_{QCD}$. If the generator of SRG transformations 
does not include interactions in $G_\lambda$, the effective 
coupling constant in the model quickly increases to 
very large values as soon as the SRG scale parameter 
$\lambda$ is lowered down to the momentum scale that 
characterizes formation of a bound state. This scale 
is much larger than the scale associated with confinement, 
which concerns distances far beyond the size of a single 
hadron. If the generator of SRG transformations includes 
interactions in $G_\lambda$, the SRG parameter $\lambda$ 
can be brought down to the momentum scale that characterizes 
the bound state and $g_\lambda$ does not increase to large 
values. The SRG procedure also helps us recognize a connection 
between asymptotic freedom and limit cycle, and the model 
example shows that in order to handle the case of limit 
cycle the generator of SRG transformations must include 
interactions in $G_\lambda$.

\vskip.2in
\centerline{\bf Acknowledgement}
\vskip.1in
The author would like to thank Jean-Fran\c cois
Mathiot and the Organizing Committee of LC2008
for invitation and outstanding hospitality.

\end{document}